\documentclass[fleqn,preprint,5p]{elsarticle}

\usepackage{hyperref}
\usepackage{color}

\usepackage{amsmath, amssymb, amsfonts}
\usepackage{graphicx}
\usepackage{hyperref}

\journal{Physics Letters B}
\usepackage{epsfig} 
\usepackage{cancel}
\newcommand{\be}{\begin{equation}}
\newcommand{\ee}{\end{equation}}
\newcommand{\bes}{\begin{equation*}}
\newcommand{\ees}{\end{equation*}}
\newcommand{\bea}{\begin{eqnarray}}
\newcommand{\eea}{\end{eqnarray}}
\newcommand{\beas}{\begin{eqnarray*}}
\newcommand{\eeas}{\end{eqnarray*}}
\newcommand{\Tr}{\text{Tr}}

\newcommand{\newc}{\newcommand}
\newc{\bi}{\begin{itemize}}
\newc{\ei}{\end{itemize}}

\newc{\ra}{\rightarrow}
\newc{\sq}   {\mbox{$\wt{q}$}}
\newc{\msq}  {\mbox{$m_{\sq}$}}
\newc{\gl}   {\mbox{$\wt{g}$}}
\newc{\mgl}  {\mbox{$m_{\gl}$}}

\newc{\wt}{\widetilde}

\newc{\ifb}{\mbox{${\rm fb}^{-1}$}}

\newc{\del}{\delta}

\begin{document}

\begin{frontmatter}

\title{Theoretical constraints on  masses of heavy particles \\ in Left-Right Symmetric Models}

\author[a]{J.~Chakrabortty} 
\author[b]{J.~Gluza}
\author[b]{ T.~Jeli\'nski}
\author[a]{T.~Srivastava} 
\address[a]{Department of Physics, Indian Institute of Technology, Kanpur-208016, India} 
\address[b]{Institute of Physics, University of Silesia, Uniwersytecka 4, 40-007 Katowice, Poland}

\begin{abstract}
Left-Right symmetric models with general $g_L \neq g_R$ gauge couplings which include bidoublet and triplet scalar multiplets are studied. Possible scalar mass spectra are outlined by imposing Tree-Unitarity, and Vacuum Stability criteria and also using the bounds on neutral scalar masses $M_{\rm H^{ FCNC}}$ which assure  the absence of Flavour Changing Neutral Currents (FCNC).  We are focusing on mass spectra relevant for the LHC analysis, i.e., 
the  scalar masses are around TeV scale.  As all non-standard heavy particle masses are related to the vacuum expectation value (VEV) of the right-handed triplet ($v_R$), the combined effects of relevant Higgs potential parameters  and $M_{\rm H^{ FCNC}}$  regulate the lower limits of heavy gauge boson masses. The complete set of Renormalization Group Evolutions for all couplings  are provided at the 1-loop level, including the mixing effects in the Yukawa sector. Most of the scalar couplings suffer from the Landau poles at the intermediate scale $Q \sim 10^{6.5}$ GeV, which in general coincides with violation of the Tree-Unitarity bounds. 
\end{abstract}


\end{frontmatter}

\section{Introduction}
\noindent
After the 2012 discovery of the spin-zero boson at the LHC \cite{Aad:2012tfa,Chatrchyan:2012ufa} we are even more convinced that the
 theoretical concept of the mass generation within the  gauge theory is correct.   The discovered particle fits well within 
 the predictions of the Standard Model (SM)  of electroweak interactions. 
In the SM the mass of the Higgs boson is a free parameter. This, along with (very) weak interaction  of the Higgs boson 
were the main reasons why it took decades to fix its mass experimentally, happened to be at the 125 GeV level \cite{Aad:2012tfa,Chatrchyan:2012ufa}.   
In the meantime many theoretical concepts connected with both the scalar sector of SM 
and perturbation techniques have been developed and understood.
It has been noted that the SM Higgs boson's  mass can be bounded from both ends using quantum field theoretical (QFT) techniques \cite{Ellis:1975ap, Veltman:1976rt, Weinberg:1976pe, Lee:1977yc, Lee:1977eg}.
These concepts are basic and general, and can be useful also nowadays when, after the LHC discovery, we would like to know much more. For instance,  what is the actual representation of the scalar multiplets and
what is the shape of the scalar potential of the fundamental theory in particle physics?
A priori, the SM theory is not the end of the story, for many reasons. 

One of the main theoretical constraints on the SM Higgs boson mass comes from the simple fact that its mass depends on the strength of the Higgs quartic coupling, so the mass 
should not exceed an upper limit above which the theory is strongly coupled and in turn the perturbative QFT is invalid. In other words, to have a consistent weakly coupled theory involving the Higgs boson, its mass must be smaller than that upper limit.
 This  constraint of weak interactions at high energies is called the unitarity limit.
    In the context of SM, the upper limit of the SM Higgs boson mass must be within $\mathcal{O}(G_F^{-1/2})$ as deduced long time ago \cite{Ellis:1975ap, Veltman:1976rt, Weinberg:1976pe, Lee:1977yc, Lee:1977eg}. This  limit had been computed more precisely in \cite{Lee:1977yc, Lee:1977eg} as $\sqrt{8\pi \sqrt{2}/3}G_F^{-1/2} \simeq \mathcal{O}(\mathrm{TeV})$.

 This is very important to understand the weakly coupled limit of all beyond Standard Model (BSM) theories which are considered, and which are tested in present  or future accelerators, notably at the LHC. The problem has been already worked out within some popular and basic models involving two Higgs doublet models (THDM) \cite{Huffel:1980sk,Kanemura:1993hm,Akeroyd:2000wc,Horejsi:2005da,Chakraborty:2014oma,Chakrabarty:2015kmt,
 Chakrabarty:2016smc} or models involving triplet scalar multiplets \cite{Aoki:2007ah,Dey:2008jm}.
Unitarity constraints have been considered in \cite{Mondal:2015fja} in the context of the  Minimal Left-Right Symmetric model (MLRSM) which contains an enriched Higgs sector:
a bidoublet and two triplets  scalar fields
\cite{Senjanovic:1975rk, Mohapatra:1979ia,Mohapatra:1980yp}.  
Some basic remarks on unitarity in the scalar sector of MLRSM can be also found in the seminal work \cite{Gunion:1989in}. In a recent paper \cite{Maiezza:2016bzp} 
perturbativity and mass scales of Left-Right Higgs bosons are also discussed.
 
In the present study we derive Tree-Unitarity (TU) constraints in MLRSM which are  written in form of  individual and (or) linear combinations of the quartic couplings. Thus these bounds are easily translated in terms of the physical scalar masses. We have also combined the Vacuum Stability (VS) criteria (for recent work on this subject in a general context, see \cite{Kannike:2016fmd})
and TU constraints with Flavour Changing Neutral Currents (FCNC) bounds which give an additional limit on the mass of the right-handed charged gauge boson. In addition, in the present work we have come up with a complete set of renormalization group equations (RGEs) 
and perform the necessary RGEs of quartic couplings. 
 
Interestingly, the concept of Left-Right (LR) symmetry has been revived recently at the LHC in the context of  dilepton \cite{Deppisch:2014qpa,Deppisch:2015cua,Dobrescu:2015qna, Gluza:2015goa,Dhuria:2015swa}, diboson  \cite{Brehmer:2015cia,Dev:2015pga} and diphoton \cite{Chakrabortty:2015hff,Dey:2015bur,Dasgupta:2015pbr,Dev:2015vjd,Deppisch:2016scs,Borah:2016uoi,
Hati:2016thk,Ren:2016gyg,Huong:2016kpa} excesses, which  might be connected with heavy particles of LR models. It is then useful  to understand possible contributions to such  signals coming from the scalar sector of the theory in future studies (first results can be found in \cite{Dobrescu:2015yba}) using bounds on the scalar sector of the theory. In the context of MLRSM we started such analysis in \cite{Chakrabortty:2012pp}, taking into account interplay of the collider signals with low energy precision data. In that paper we treated Higgs boson masses practically as free parameters, not taking into account many possible theoretical constraints. Nonetheless, there we showed  that correlations between the Higgs bosons and gauge bosons as well as the radiative muon decay at 1-loop level  impose strong constraints on high energy LHC signals.   
To understand the realistic scalar spectrum of the theory, dedicated analyses have been further performed  in \cite{Bambhaniya:2013wza, Bambhaniya:2014cia, Bambhaniya:2015wna, Dev:2016dja}. In these papers the constraints from FCNC, VS along with the LHC exclusions were considered.
   It has been found among others that not all four charged Higgs bosons of the theory can be simultaneously light (below 1 TeV).  Taking into account this limitation, we have found several benchmark points \cite{Bambhaniya:2014cia, Bambhaniya:2015wna} which are within reach of the LHC future runs. 
For other studies of the Higgs sector of the theory, see e.g. \cite{Senjanovic:1978ee,Grifols:1978wk,Olness:1985bg,Frank:1991sy,Chang:1992bg,Maalampi:1993tj,Gluza:1994ad,Bhattacharyya:1995nt,Boyarkina:2000bn,Barenboim:2001vu,Gogoladze:2003bb,Azuelos:2004mwa,Kiers:2005gh,Jung:2008pz,Guadagnoli:2010sd,Blanke:2011ry,Mohapatra:2013cia,Aydemir:2014ama,Maiezza:2015lza,Maiezza:2015qbd,Dev:2016dja}.

Here, we incorporate TU constraints and further extend the analysis. We also take care of the constraints and potentially problematic structures due to Landau poles which arise from the concept of RGEs \cite{ZinnJustin:1996cy}.  
 RGEs in MLRSM have been considered at the one-loop level, originally in  \cite{Rothstein:1990qx}. Here, we have performed independent  RG analysis after correcting some misprints in the published article, see Sec.~\ref{rges} of the present work for details. In addition, we provide a complete set of 1-loop RGEs, including all couplings of the theory. It is important for two reasons:  (i) to prepare a well-tested background for higher-loops analysis, and (ii) the earlier results \cite{Rothstein:1990qx} have been used repeatedly in recent studies \cite{Chakrabortty:2013mha,Chakrabortty:2013zja,Mondal:2015fja} 
and it is better  to avoid proliferation of misprints in the future. 

In the SM, as the EWSB scale is determined from the observed gauge boson masses, the upper limit on the SM Higgs boson can be fixed. Similarly, if in a near future the right-handed gauge boson masses are fixed from observation then the absolute upper mass  bounds of the scalars can be provided. Thus, as of now, the bounds depend on the $SU(2)_R$  breaking scale $v_R$. In this paper upper limits on 
the heaviest mass of these Higgs bosons compatible with the TU bounds are computed as  functions of $v_R$.  

\section{Model: Left-Right Symmetry \label{LR}}

The model is based on the  $SU(2)_L \otimes SU(2)_R \otimes U(1)_{B-L}$ 
Left-Right gauge symmetry (LR) \cite{Senjanovic:1975rk, Mohapatra:1979ia,Mohapatra:1980yp}. The spontaneous symmetry breaking occurs in two steps: $SU(2)_R \otimes U(1)_{B-L} \to U(1)_Y$, and $SU(2)_L \otimes U(1)_Y \to U(1)_{em}$. 
To achieve this symmetry breaking we choose a traditional spectrum of Higgs sector multiplets
with a bidoublet and two triplets \cite{Mohapatra:1980yp,Gunion:1989in}
\begin{equation}
\phi =\left( 
\begin{array}{lr}
\phi_1^0 \;&\; \phi_1^+\\
\phi_2^- & \phi_2^0
\end{array} 
\right)\equiv [2,2,0], 
\end{equation}  
\begin{equation}  
\Delta_{L(R)}=\left( 
\begin{array}{cc}
\delta_{L(R)}^+/\sqrt{2} & \delta_{L(R)}^{++}\\     
\delta_{L(R)}^0 & -\delta_{L(R)}^+/\sqrt{2}
\end{array} 
\right) \equiv [3(1),1(3),2],
\end{equation}
\\
where the quantum numbers in square brackets are given for 
$SU(2)_L$, $SU(2)_R$ and  $U(1)_{B-L}$ groups, respectively. 

The vacuum expectation values (VEVs) of the scalar fields can be recast in the following form:
\begin{equation}
\left< \phi \right>  =\left( \begin{array}{cc}
                        \kappa_1/\sqrt{2} & 0\\
                        0      \;&\; \kappa_2/\sqrt{2}
                       \end{array}\right) , \hskip 2pt                       
\left< \Delta_{L,R} \right>  =\left( \begin{array}{lr}
                        0   & 0\\
                        v_{L,R}/\sqrt{2}\; & \;0
                       \end{array}\right).                       
\label{vev}
\end{equation}

VEVs of the right-handed triplet ($\Delta_R$) and the bi-doublet ($\phi$), propel the respective symmetry breaking:  \\
$SU(2)_R \otimes U(1)_{B-L}\to U(1)_Y$, and $SU(2)_L \otimes U(1)_Y \to U(1)_{em}$. As $v_L\ll \kappa_{1,2}\ll v_R$, we take safely $v_L=0$.

We set  the coefficients of the quartic couplings that are linear in $\Delta_{L,R}$ to be zero \cite{Deshpande:1990ip}. 
We also assume that the right-handed symmetry breaking scale, $v_R$, is much larger than the electroweak scale, $\kappa_+ \equiv \sqrt{\kappa_1^2 + \kappa_2^2}$. Thus the terms proportional to $\kappa_+$ will be neglected comparing to the terms proportional to $v_R$.
This assumption is phenomenologically viable and supported also by the exclusion 
limits given by the LHC. In addition $\kappa_1 \gg \kappa_2 \simeq 0$ \cite{Deshpande:1990ip}. These relations simplify correlations among the unphysical and physical Higgs fields which are related to each other by Eq.~(74) in \cite{Duka:1999uc}.

\section{Unitarity bounds \label{UB}}
\noindent
The quartic part of the scalar potential can be written in terms of the physical fields as follows:
\begin{equation}
 V(H_{0,1,2,3}^0;A_{1,2}^0 ; H_{1,2}^\pm ; H_{1,2}^{\pm \pm}) =  
\sum_{m=1,..,72} \Lambda_m H_i H_j H_k H_L, \nonumber
\end{equation}
where $H_i,H_j,H_k, H_l \in ( H_{0,1,2,3}^0;A_{1,2}^0 ; H_{1,2}^\pm ; H_{1,2}^{\pm \pm} )$.
To understand the unitarity constraints one needs to look at the following scattering processes \cite{Huffel:1980sk}:
\begin{equation}
H_i + H_j \to H_p + H_q,
\end{equation}
where $H_{i,j,p,q}$ are the physical Higgs fields. These scatterings can happen in two ways at the tree level through: (i) Contact terms, i.e., four point scalar couplings which are outcome of the scalar quartic potential, and (ii) Higgs-Higgs-Gauge boson couplings. We know that the Higgs-Higgs-Gauge boson couplings contain derivatives 
owing to their Lorentz structure, thus when they are connected with the gauge boson exchange diagrams the maximum divergences which can  appear through these diagrams are logarithmic. Considering theories up to the Planck scale, we do not need to worry about the logarithmic unitarity violations  \cite{Huffel:1980sk}.

One can estimate the strength of these scalar four-point contact interactions in two ways. First, consider the process in terms of the unphysical scalar fields and reconstruct all the elements in terms of the physical neutral and charged scalars. In this case a vertex factor will be a polynomial function of the couplings which can be thought of as a rotated quartic coupling basis. As the model under consideration contains many scalar field components, it would be difficult to pin down the unitarity bounds in terms of the couplings and translate them to the masses of the scalar fields.  There is an alternative option which we have adopted in this paper. Instead of rotating the quartic couplings we have sorted  out all possible quartic contact terms in terms of the physical fields where the vertex factors of each coupling are linear functions of the quartic couplings. In this way, we can immediately find out the unitarity bounds on the quartic couplings. This is also helpful to translate the bounds in terms of the masses of the physical scalar fields as the mass terms posses linear dependence on the quartic couplings. Thus the unitarity bounds on the scalar masses can be easily incorporated, which is our prime aim in this analysis.\\
  Our further strategy is as follows:  
to invoke that the scalars are weakly coupled we must satisfy the inequality: $|\Lambda_m| < 8 \pi$ \cite{Marciano:1989ns} for the scalar quartic couplings. There are many of them, so these couplings are gathered in  \cite{files}. Let us note that this is an improvement over \cite{Huffel:1980sk} where the unitarity  bound was given as $|\Lambda_m| < 16 \pi$. For the sake of analysis 
it is sufficient to identify the  couplings with the largest coefficients.
For example, if coupling $\lambda_i$ appears with coefficient $a_1$ and $a_2$ such that $a_1>a_2$, then for the unitarity constraint the term  $a_1\lambda_i$ is considered, the second term  will respect the unitarity bound on $\lambda_i$ automatically. 
  
As all  terms with quartic couplings in four-scalar scatterings must be smaller than $8 \pi$, the following   constraints on the quartic couplings follow:
\begin{eqnarray}
&&\lambda_1 < 4 \pi/3,~ (\lambda_1+4 \lambda_2+2 \lambda_3) < 4 \pi,\label{lsTU1}\\
&&(\lambda_1-4 \lambda_2+2 \lambda_3) < 4 \pi,\label{lsTU2}\\ 
&&\lambda_4 < 4 \pi/3, \label{treeuni1}\\
&&\alpha_1 < 8 \pi, ~\alpha_2 < 4 \pi,~ (\alpha_1+\alpha_3) < 8 \pi, \label{treeuni2}\\
&&\rho_1 < 4 \pi/3, ~( \rho_1+\rho_2) < 2 \pi,~ \rho_2< 2\sqrt{2}\pi,\label{rsTU1}\\
&&\rho_3 < 8\pi, ~\rho_4 < 2\sqrt{2}\pi. \label{treeuni3}
\end{eqnarray}
The scalar spectrum is\footnote{In Eq.~(\ref{vev}) VEVs are normalized by $\sqrt{2}$, so $\kappa_+=246$ GeV, as in  \cite{Gunion:1989in,Deshpande:1990ip,Duka:1999uc}. In \cite{Maiezza:2016bzp,Dev:2016dja} there is no such VEVs normalization, so $\kappa_+=174$ GeV, and the mass relation for the SM equivalent Higgs boson $H_0^0$ in Eq.~(\ref{spec1}) differs accordingly.}:
\begin{eqnarray}
M_{H_0^0}^2 & = & 2\left(\lambda_1-\frac{\alpha_1^2}{4\rho_1}\right) \kappa_+^2,\label{spec1}\\
M_{H_1^0}^2 & = & \frac{1}{2} \alpha_3 v_R^2 <  4 \pi v_R^2,\label{MH10sq}\\
M_{H_2^0}^2 & = & 2 \rho_1 v_R^2  < (8 \pi/3) v_R^2,\\
M_{H_3^0}^2 & = & \frac{1}{2} (\rho_3 -2 \rho_1) v_R^2 < (4\pi v_R^2-M_{H_2^0}^2/2), \\
M_{A_1^0}^2 & = & \frac{1}{2} \alpha_3 v_R^2 -2 \kappa_+^2 (2\lambda_2 - \lambda_3) < 4 \pi v_R^2,\label{MA10sq}\\
M_{A_2^0}^2 & = & \frac{1}{2} (\rho_3 -2 \rho_1) v_R^2 < (4\pi v_R^2-M_{H_2^0}^2/2), 
\end{eqnarray}
\begin{eqnarray}
M_{H_1^{\pm}}^2 & = & \frac{1}{2} (\rho_3 -2 \rho_1) v_R^2 + \frac{1}{4} \alpha_3 k_+^2\nonumber\\
&<& (4\pi v_R^2-M_{H_2^0}^2/2), \\
M_{H_2^{\pm}}^2 & = & \frac{1}{2} \alpha_3 v_R^2 +\frac{1}{4}\alpha_3 k_+^2 < 4 \pi v_R^2,\\
M_{H_1^{\pm \pm}}^2 & = &  \frac{1}{2}(\rho_3 -2 \rho_1) v_R^2 + \frac{1}{2}\alpha_3 k_+^2 \nonumber\\
&<& (4\pi v_R^2-M_{H_2^0}^2/2),\\
M_{H_2^{\pm \pm}}^2 & = & 2 \rho_2 v_R^2 + \frac{1}{2}  \alpha_3 k_+^2 <
 4\sqrt{2} \pi v_R^2.\label{MH2pp2}
\end{eqnarray}

In  \cite{Duka:1999uc} the second term in Eq.~\ref{spec1} has been missed and we sketch its derivation in the Appendix. 
After the Higgs boson discovery, this mass relation is fixed and can be helpful  for RGEs discussion, see Section~\ref{rges}.

\section{Vacuum Stability Criteria}
\noindent
Apart from the TU constraints discussed in the previous section, the quartic couplings have to satisfy necessary conditions for the vacuum stability \cite{Chakrabortty:2013zja, Chakrabortty:2013mha}:
\begin{eqnarray}\label{stability}
\lambda_{1} \geq 0,~~ \rho_1 \geq 0,~~ \rho_1+ \rho_2 \geq 0,~~ \rho_1+ 2\rho_2 \geq 0.
\end{eqnarray}

In passing we would like to emphasize few comments on computation of vacuum stability criteria.
We have used the vacuum stability criteria computed in \cite{Chakrabortty:2013mha} using the copositivity conditions, which is an improved version of the positivity idea used in \cite{Chakrabortty:2013zja}. 
The copositivity criteria lead to the vacuum stability conditions  which encapsulate broader parameter space than that  comes from  the positivity criteria
\cite{Chakrabortty:2013zja}.  Thus, it is indeed possible that for some values of quartic couplings the vacuum looks to be unbounded from below, if we use former criteria. In reality that may not be true, if they satisfy copositivity criteria. Thus, the copositivity criteria as computed in \cite{Chakrabortty:2013mha} are certainly an improvement over results given in \cite{Chakrabortty:2013zja}. Here, we would like to mention that one must be careful while computing the copositivity criteria as it has some basis dependency, and for some choices of basis it is possible to encounter some unrealistic stringent criteria.

\begin{figure}[htb]
\begin{center}
\includegraphics[scale=0.65]{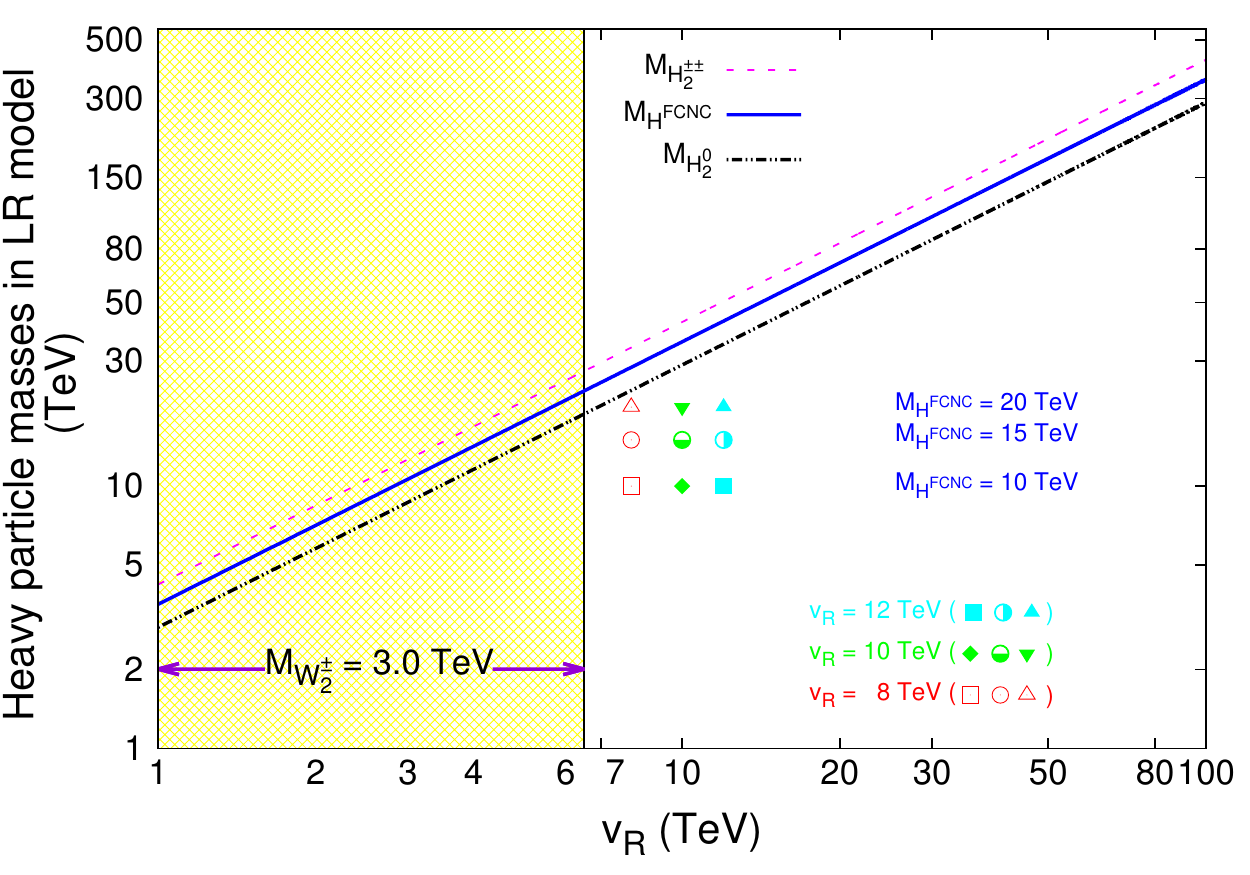}
\caption{\label{fig1} Upper limits on masses of scalars in MLRSM as a function of $v_R$. In this plot we define: 
${\rm M_{H^{\rm FCNC}}} \in [M_{H_{1}^{0}},M_{A_{1}^{0}}]$; $M_{H_0^0}$ is the SM Higgs. These limits are outcome of unitarity and vacuum stability constraints discussed in the text. 
The shaded region of $v_R$ is due to the exclusion limits on $W_2^{\pm}$ experimental searches, which give typically 3.0 TeV \cite{Khachatryan:2014dka}  for the restricted MLRSM scenario. Three benchmark points discussed in \cite{Bambhaniya:2014cia}, corresponding to $v_R=8$ TeV: ${\rm M_{H^{\rm FCNC}}}$=10 (box), 15 (circle), 20 (triangle) TeV are shown. They are compatible with low energy constraints, and also TU and VS constraints for that particular choice of $v_R$.}
\end{center}
\end{figure}

From Eqs.~(\ref{lsTU1})-(\ref{MH2pp2}) it is easy to note that it is not possible to compute the upper limits on the masses of all the scalars individually. 
This is because for some of the quartic couplings the unitarity constraints  are quite entangled and cannot be decoupled. Thus, the upper limits of a few scalar masses are functions of masses of other scalars, e.g., maximum values of $M_{H_3^0},M_{A_2^0},M_{H_1^{\pm}},M_{H_1^{\pm\pm}}$ depend on $M_{H_2^0}$. So,  the unitarity constraints on their masses do not lead to upper limits. Among all the scalars, $H_2^{\pm \pm}$ can be the heaviest for all choices of $v_R$, see Fig.~\ref{fig1}. For $M_{H_1^0}$ and $M_{A_1^0}$ the vacuum stability criteria allows to set the mass upper limits, which would not be possible if we used only unitarity bounds. In Fig.~\ref{fig1} the upper limits on $M_{H_1^0},M_{A_1^0}$ respect the vacuum stability as well as unitarity constraints. 
In Fig.~\ref{fig1} some benchmark points (BP) discussed in \cite{Bambhaniya:2014cia} are included, for readers convenience, the exact spectrum and scalar potential parameters for $v_R=12$ TeV are repeated in the Appendix. These BPs lead to the degenerate doubly charged Higgs bosons within reach of the LHC  and also satisfy VS and FCNC criteria, see Eqs.(9)-(13) in \cite{Bambhaniya:2014cia}.
As we can see, all of them are at the allowed region, though BPs for $M_{\rm H^{ FCNC}}=20$ TeV marginally (larger masses are disfavoured).

Let us discuss the limit on  mass of the gauge boson $W_2$ related to TU and VS of the scalar potential. 


\begin{figure}[htb]
\begin{center}
\includegraphics[scale=0.65]{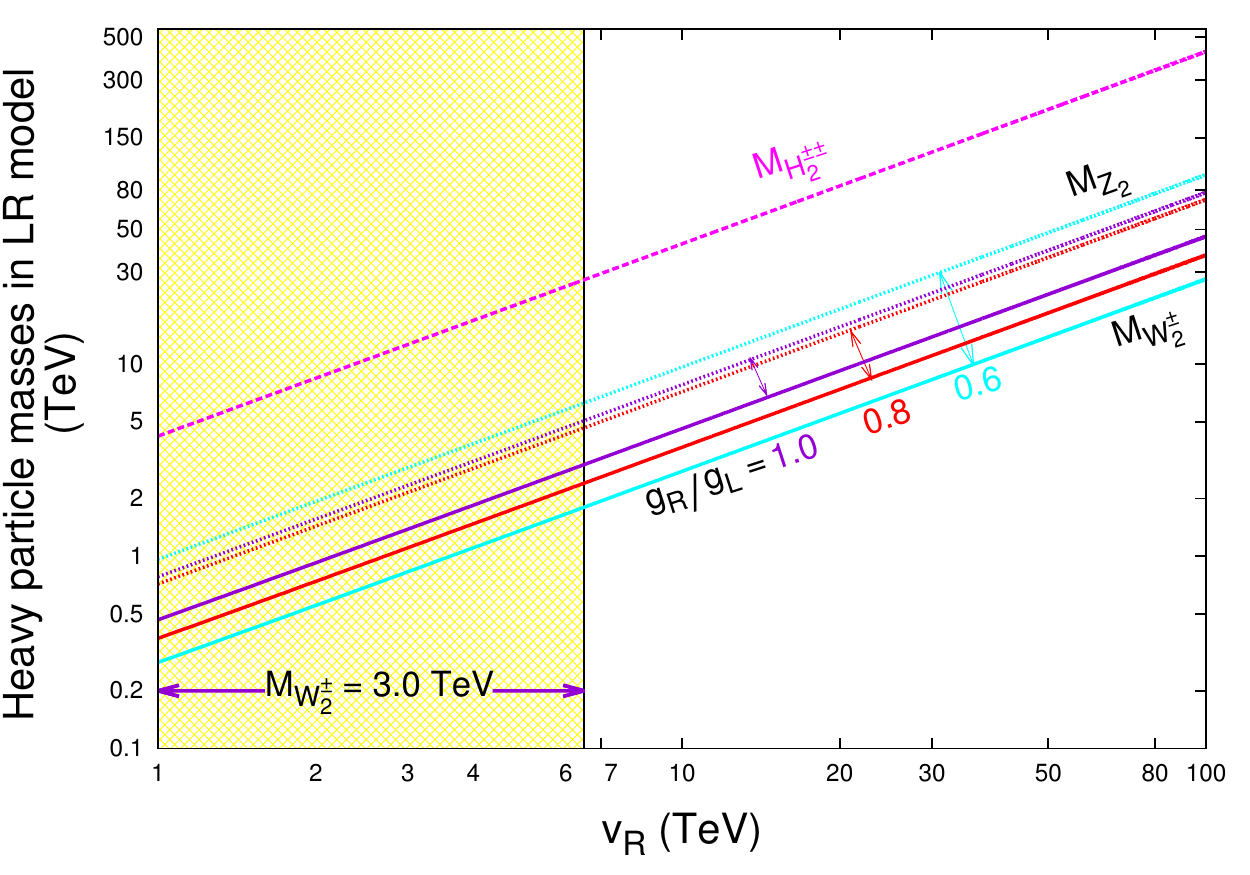}
\caption{\label{fig3} Masses of heavy gauge bosons in $\cancel{\it{M}}$LRSM scenarios. The mass splittings among $W_2^{\pm}$ and $Z_2$ are shown for 
$g_R/g_L$=1.0, 0.8, 0.6.
The shaded region of $v_R$ is as in Fig.1. For comparison, the upper limit on $H_2^{\pm \pm}$ from Fig.~\ref{fig1} is included here. 
}
\end{center}
\end{figure}

We consider the minimal version (MLRSM) of the left-right model where the gauge couplings are equal $g_{L}=g_R$, and its non-minimal version ($\cancel{\it{M}}$LRSM) where $g_L \neq g_R$. The latter scenario seems to be more suitable if strict gauge coupling unification is assumed 
 \cite{Chakrabortty:2009xm}.  
This choice was also discussed in the context of the   LHC diboson excesses  in \cite{Deppisch:2014qpa,Gluza:2015goa,Brehmer:2015cia,Dhuria:2015swa}.
In $\cancel{\it{M}}$LRSM scenario the gauge boson masses are given in an analytical form as \cite{Jelinski:2015ifw} ($g_a=g_R/g_L$):
\begin{eqnarray}
M_{W_2}^2 & = & \frac{g_L^2}{8} \Bigg[ (1+g_a^2) \kappa_+^2+2 g_a^2 v_R^2\nonumber\\
&&+\sqrt{16 g_a^2 \kappa_1^2 \kappa_2^2+((g_a^2-1) \kappa_+^2+2 g_a^2 v_R^2)^2}\Bigg], \\
M_{Z_2}^2 & =&\frac{1}{8} \left\lbrace 4 g^{'2}  v_R^2\right.\nonumber\\
&&+g_L^2 v_R^2 \left[\left(4 g_{a}^2+\frac{4 g^{'2}}{g_L^2}+\frac{\left(1+g_{a}^2\right) \kappa_+^2}{v_R^2}\right)^2\right.\nonumber\\
&&\left.-\frac{16 \left(g^{'2}+g_{a}^2 \left(g_L^2+g^{'2}\right)\right) \kappa_+^2}{g_L^2 v_R^2}\right]^{1/2}\nonumber \nonumber\\
&&\left.+g_L^2 \left(\kappa_+^2+g_{a}^2 \left(\kappa_+^2+4 v_R^2\right)\right) \right\rbrace.
\end{eqnarray}
 $g^{'}$ is the gauge coupling corresponding to $U(1)_{B-L}$ gauge symmetry.
 In Fig.~\ref{fig3} masses of heavy gauge bosons are given. They depend on gauge couplings and the mass splitting between charged and neutral gauge bosons increases with decreasing $g_R/g_L$ ratio. 
  This is quite clear if we look at the correlations among the gauge couplings. As $g_R$ decreases one needs larger value of $g'$ to ensure proper value of $U(1)_Y$ gauge coupling, $g_Y$. That in turn increases $M_{Z_2}$, and thus the splitting is enlarged.
 Let us note that naturally $M_{Z_2} > M_{W_2}$, for more exotic scenarios, see \cite{Patra:2015bga}.

As $\kappa_+\ll v_R$,   mass of $W_2$ in MLRSM ($g_L=g_R=g_2$) is given by $M_{W_2}=g_2v_R/\sqrt{2}$. Hence, a limit on $M_{W_2}$ is strictly related to the limit on $v_R$. The latter, in turn, has to be bigger than
\begin{eqnarray}\label{vRFCNC}
v_R\gtrsim \frac{\sqrt{2}M_{\rm H^{FCNC}}}{\sqrt{\alpha_3}},
\end{eqnarray}
in order to ensure that masses of $H_1^0$ and $A_1^0$ are greater than $M_{\rm H^{ FCNC}}\approx10\,\mathrm{TeV}$. This is the lowest limit on FCNC Higgs bosons
 \cite{Guadagnoli:2010sd}, one of the strongest limits has been obtained in  
\cite{Pospelov:1996fq} ($M_{\rm H^{ FCNC}} \geq 50 \,\mathrm{TeV}$). 
Taking $M_{\rm H^{ FCNC}}\approx 10\;(20,50)\,\mathrm{TeV}$ and $\alpha_3 \leq 8 \pi$, see Eq.~(8), we get 

\begin{equation}
M_{W_2} \geq \frac{g_2 M_{\rm H^{ FCNC}}}{\sqrt{8\pi}}\approx1.3\; (2.6,6.5) \;\mathrm{TeV}. \label{mw2}
\end{equation}
This is the lowest  limit on the charged gauge boson mass 
with a minimal theoretical assumption which takes into account scalar sector of the model.

Similar bounds as in Eq.~\ref{mw2} can be obtained for $M_{Z_2}$, $M_{Z_2} \simeq 1.66 \times M_{W_2}$ in MLRSM.

\section{Renormalization Group Evolution \label{rges}}
\noindent
In the SM, after Higgs boson discovery\footnote{The Higgs boson mass is not a free parameter any more and RGEs need one less free parameter.} there are arguments that at 1-loop and beyond there are no Landau poles up to the Planck mass scale \cite{Jegerlehner:2013cta}.  
It is interesting to note that the Higgs self-coupling $\lambda$ as well as the top-quark Yukawa coupling $y_t$ at one loop are asymptotically free for parameter range fixed by recent data. This is not changed by higher corrections up to three loops. However, if the Higgs self coupling would be bigger, there would be a Landau pole at very high scales, see for instance \cite{Hambye:1996wb}.  For SM the existence of a Landau pole depends mainly on the value of the top-quark Yukawa coupling $y_t$. 
Here, the situation is not very transparent. In some analysis, e.g. \cite{Buttazzo:2013uya}, the Higgs $\beta$-function vanishes around $10^9$ GeV, whereas according to \cite{Jegerlehner:2013cta} its zero occurs at about $10^{17}$ GeV with lower $y_t$. In this scenario the Landau pole is appearing but far beyond the Planck scale.

We can see how fragile are the results and conclusions based on renormalization group (RG) analysis in the SM. 
So, what can we expect within the  beyond SM scenario? Here, the higher order corrections are even more complicated.
But they can be crucial in some corner of the parameter space:  imagine that  either $\lambda$ or $y_t$ are adjusted such that
 Higgs beta function is positive. Then a Landau pole at some high scale may emerge, and it is quite possible that
 higher loop corrections can cause the change in sign of beta-functions. 
 In this way the stability analysis can be performed with better accuracy. 
 
Let us discuss RG evolution of the scalar potential parameters. To that end we shall use 1-loop RG equations. 
As computation of $\beta$ function coefficients is  error prone especially in a model with many couplings in the scalar potential, we have used the \textsc{PyR@TE} (v1.2.2 beta) package \cite{Lyonnet:2013dna,Lyonnet:2015jca} to automatically generate 1-loop 
RGEs for  MLRSM.

The explicit form of RGEs has already been given e.g. in \cite{Rothstein:1990qx}. Those formulas were latter used, e.g., in \cite{Chakrabortty:2013mha,Chakrabortty:2013zja,Mondal:2015fja}.
In this article we have made following progress and improvements in this context:
\begin{itemize}

\item
We have computed the full set of RGEs.
For example the renormalization group evolution of one of the quartic couplings is computed as: 
\begin{align*}
					 (4\pi)^2 \frac{d\lambda_{3}}{d\ln Q} =
				&+\frac{64}{3} \lambda_{2} \lambda_{3} 
			 - 9 g_{R}^{2} \lambda_{3} 	+\frac{64}{3} \lambda_{2}^{2}
			 - 9 g_{L}^{2} \lambda_{3} \\
				&	+32 \lambda_{3}^{2} 
					+24 \lambda_{1} \lambda_{3} 
			 + \mathcal{O}(\lambda y^2) + \mathcal{O}(y^4).
			\end{align*}
Let us note that there is no term $\propto g^2$ unlike given in \cite{Rothstein:1990qx}. Also a term like (+$\frac{64}{3} \lambda_{2} \lambda_{3}$) was absent in \cite{Rothstein:1990qx}.

\item We have also provided the evolutions of scalar mass parameters.

\item  Our RGEs contain right and left-handed gauge couplings separately thus can be used for non-minimal models where $g_L\neq g_R$.
\item 
The Yukawa couplings are appearing as matrices so the scenario with non-diagonal Yukawa couplings and their mixing effects can be adjudged.
 As RGEs of the couplings are coupled, these new and correct set of equations will be very important for future analyses. 

\end{itemize}

Rather than displaying all coupled complicated and clumsy equations, we have included them in related \\ \textsc{Mathematica} file which was automatically generated using \textsc{PyR@TE}. That file \verb|LR-RGEs-1-loop.m| together with numerical routines for solving 1-loop 
RGEs can be downloaded from  \cite{files}. In the Appendix we have encoded only one example of the RGEs to show their structures and complexities. 

We will not discuss the evolution of mass parameters $\mu_i^{2}$, see \eqref{V}, as they are not relevant for our analysis.
 Let us only note that
their values at the scale $v_R$ are given by extremization conditions of the scalar potential, see  \cite{Gunion:1989in,Deshpande:1990ip}. Hence $\mu_i^2(v_R)$ can be expressed with the help of initial values for remaining free parameters of the model i.e. $\alpha_i(v_R)$, $\lambda_j(v_R)$, $\rho_k(v_R)$ and mass scales $\kappa_+$, $v_R$.

For the simplicity, we assume that $v_R\sim 14\,\mathrm{TeV}$ which is safe as we have noted in our earlier section, and all the masses \eqref{MH10sq}-\eqref{MH2pp2} are $\mathcal{O}(v_R)$. 
The only mass which is fixed is the mass of the lightest Higgs boson $H_0^0$. It gives relation between values of $\lambda_1$, $\alpha_1$ and $\rho_1$, see \eqref{spec1}.  

\begin{figure}[t!]
\begin{center}
\includegraphics[width=0.88\columnwidth]{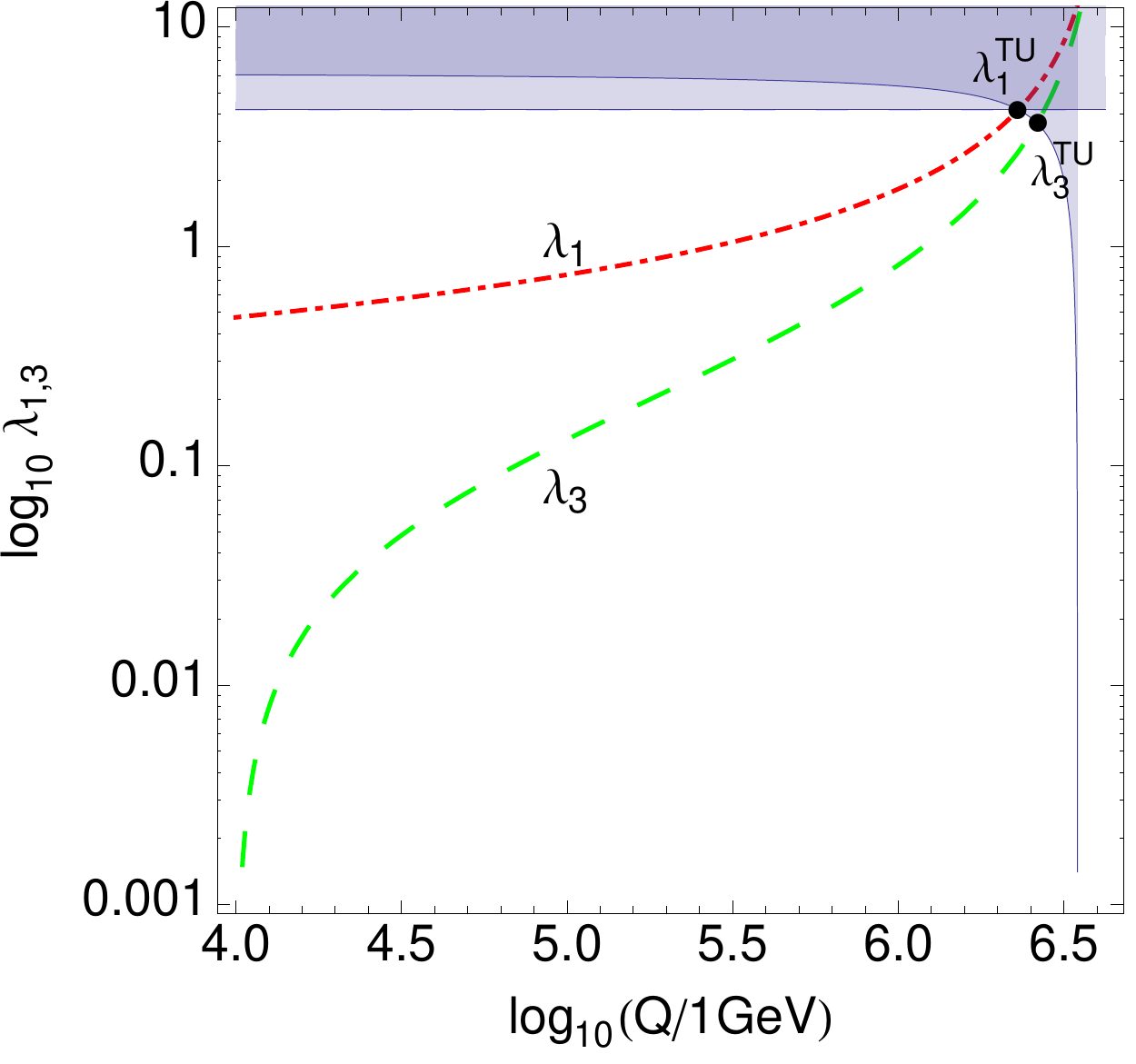}
\caption{\label{ls-RGE}
RG running of $\lambda_{1,3}$ from scale $Q=v_R$ up to $Q\approx10^{6.5}\,\mathrm{GeV}$ where Landau pole appears. 
Red dot-dashed line corresponds to $\lambda_1(Q)$ while green dashed line represents $\lambda_3(Q)$. $\lambda_{2,4}(Q)\approx0$ are not shown on the plot. Shaded region corresponds the exclusion limits provided by the  unitarity bounds \eqref{lsTU1} and \eqref{lsTU2} which need to be respected by $\lambda_1$ and $\lambda_3$. Black dots with labels $\lambda_{1,3}^{\mathrm{TU}}$ show where $\lambda_{1,3}$ enter region forbidden by the Tree-Unitarity.}
\end{center}
\end{figure}

The parameters of the scalar potential which do not explicitly enter formulas \eqref{spec1}-\eqref{MH2pp2} are set to zero at the scale $Q_0=v_R/\sqrt{2}$:
\begin{eqnarray}
\alpha_{2}(Q_0)=\lambda_{2,3,4}(Q_0)=\rho_4(Q_0)=0.
\end{eqnarray}
It turns out that such values of $\alpha_{2}$, $\lambda_{2,4}$ and $\rho_4$ are stable under RG evolution.  
To present typical behaviour of the model under RG flow let us set the values of the remaining parameters at $Q_0$ as follows:
\begin{eqnarray}\label{IC}
&&\frac{1}{2}\alpha_3(Q_0)=2\rho_{1,2}(Q_0)=\frac{1}{2}\left[\rho_3(Q_0)-2\rho_1(Q_0)\right]\nonumber\\ 
&&=\left(\frac{g_2}{\sqrt{2}}\frac{M_{\rm H^{ FCNC}}}{M_{W_2}}\right)^2\approx0.48
\end{eqnarray}
Such choice results in nearly equal masses of $H_{1,2,3}^0$, $A_{1,2}^0$, $H_{1,2}^{\pm}$ and $H_{1,2}^{\pm\pm}$ and moreover ensures that $M_{H_1^0}\approx10\,\mathrm{TeV}$. 
The initial value of $\lambda_1$ was set to 0.48. It yields a typical behaviour of that parameter under RG evolution, see Fig~\ref{ls-RGE}.  It is interesting to note that varying $\lambda_1$ in the range $[0.1 - 1.5]$ results in a shift of a position of the Landau pole from $10^7~\mathrm{GeV}$ to $10^5~\mathrm{GeV}$. Finally, let us recall that due to \eqref{spec1}, the value of $\alpha_1(Q_0)$ is also fixed. Hence all the initial conditions are specified.  
Contrary to \cite{Rothstein:1990qx}, where the RGE running of only specific terms of scalar potential couplings were considered, in this work we choose such initial values of scalar potential parameters, see \eqref{IC}, that result in a well-defined mass spectrum. It means that all the scalar masses are positive, all the experimental bounds on scalar particles masses are satisfied, and stability and Tree-Unitarity conditions are fulfilled, see Eqs.~\eqref{stability} and \eqref{lsTU1}-\eqref{treeuni3}, respectively. 
 
\begin{figure}[h!]
\begin{center}
\includegraphics[width=0.85\columnwidth]{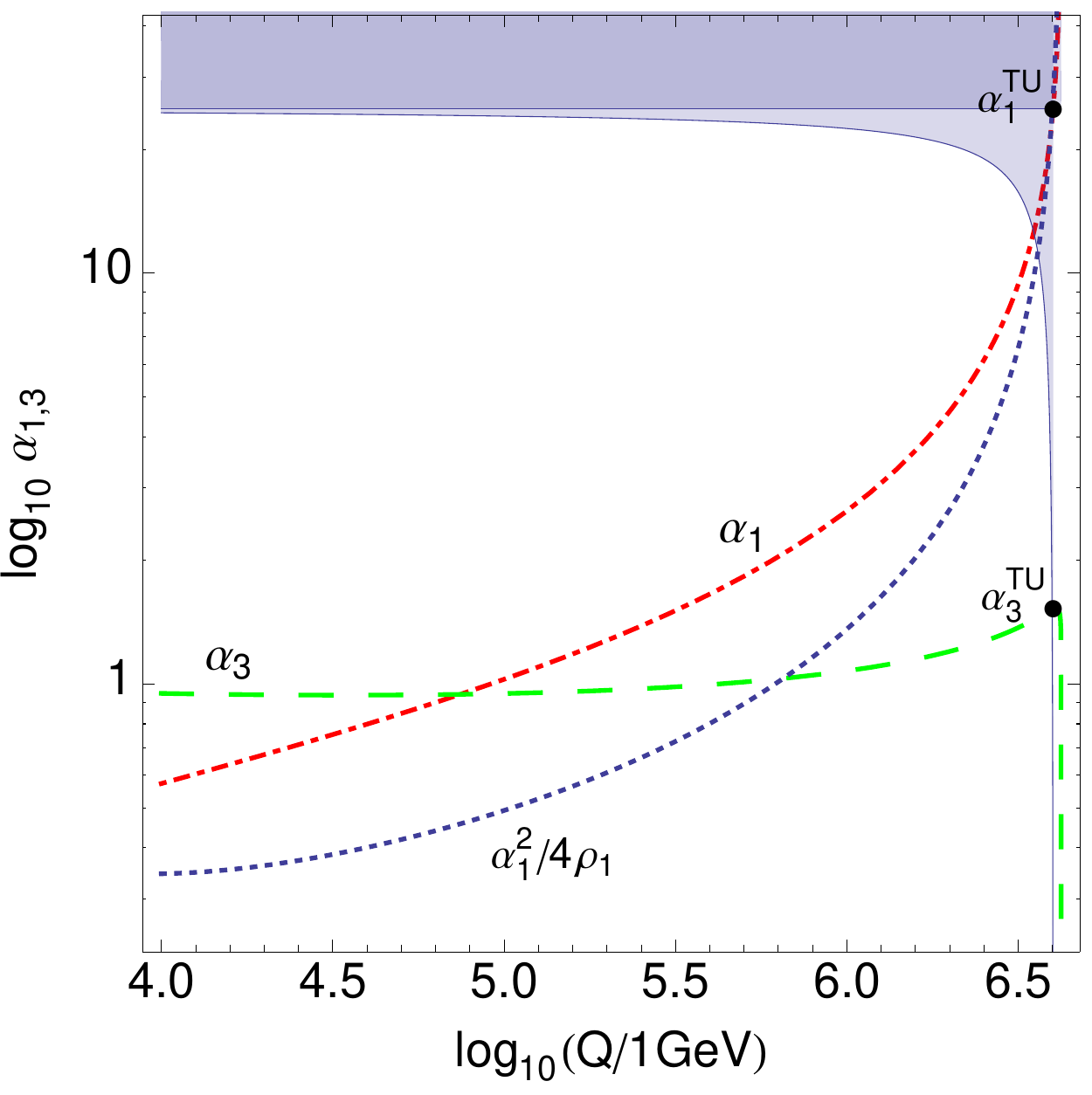}
\caption{\label{as-RGE}
RG running of $\alpha_{1,3}$ from scale $Q=v_R$ up to $Q\approx10^{6.5}\,\mathrm{GeV}$ where Landau pole appears. Red dot-dashed line corresponds to $\alpha_1(Q)$ while green dashed line represents $\alpha_3(Q)$. $\alpha_{2}(Q)\approx0$ is not shown on the plot.  Shaded region corresponds the exclusion limits provided by the  unitarity bounds \eqref{treeuni2} which must be respected by $\alpha_1$ and $\alpha_3$. Black dots with labels $\alpha_{1,3}^{\mathrm{TU}}$ show where $\alpha_{1,3}$ enter region forbidden by the Tree-Unitarity. Blue dotted line displays RG running of the ratio $\alpha_1^2/4\rho_1$ which appears in \eqref{spec1}.}
\end{center}
\end{figure}
\begin{figure}[h!]
\begin{center}
\includegraphics[width=0.85\columnwidth]{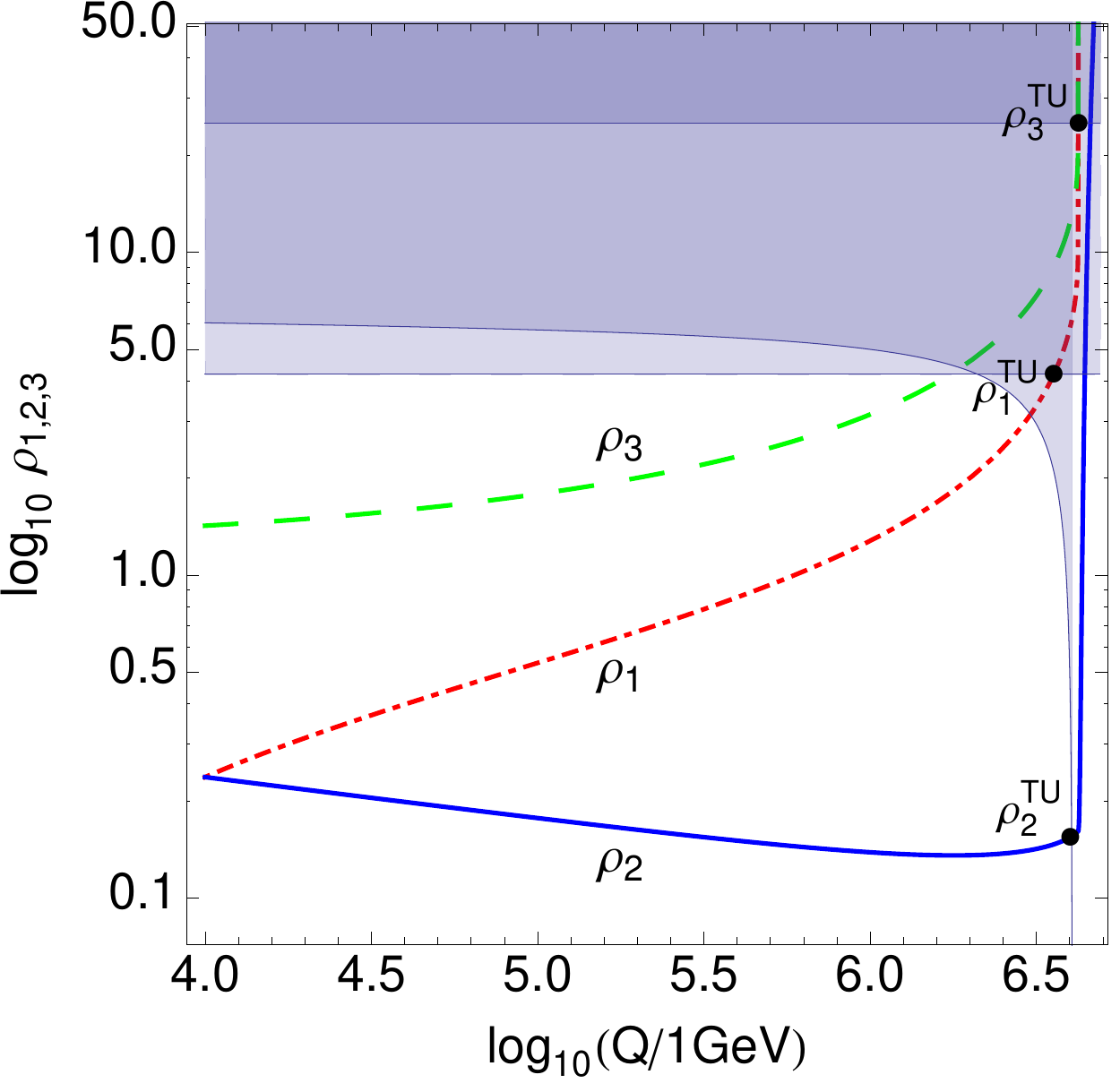}
\caption{\label{rs-RGE}
RG running of $\rho_{1,2,3}$ from scale $Q=v_R$ up to $Q\approx10^{6.5}\,\mathrm{GeV}$ where Landau pole appears. Red dot-dashed line corresponds to $\rho_1(Q)$,  blue solid line represents $\rho_2(Q)$, while green dashed line shows $\rho_3(Q)$. $\rho_{4}(Q)\approx0$ is not displayed on the plot.  Shaded region corresponds the exclusion limits provided by the  unitarity bounds \eqref{rsTU1} and \eqref{treeuni3} which need to be respected by $\rho_1$, $\rho_2$ and $\rho_3$. Black dots with labels $\rho_{1,2,3}^{\mathrm{TU}}$ show where $\rho_{1,2,3}$ enter region forbidden by the Tree-Unitarity.}
\end{center}
\end{figure}
 
The obtained RG flow of the parameters from the renormalization scale $Q_0\approx10^4\,\mathrm{GeV}$ up to higher scales is shown in Figs.~\ref{ls-RGE},~\ref{as-RGE} and \ref{rs-RGE}. Let us shortly discuss these results. First, the Landau-pole-type behaviour of $\lambda_{1,3}$, $\alpha_{1,3}$ and $\rho_{1,2,3}$ is clearly visible. The reason for this phenomenon are self-couplings of these parameters which dominate positive contributions to their $\beta$ functions. As a consequence e.g. $\lambda_{1,3}$, and similarly other couplings, start to increase rapidly when the renormalization scale $Q$ approaches \linebreak[4] $10^{6.5}~\mathrm{GeV}$ leading to Landau pole at that scale. 
 Moreover at $Q\approx10^{6.5}~\mathrm{GeV}$ the perturbative description of MLRSM breaks down due to the violation of the Tree-Unitarity bounds \eqref{treeuni1}-\eqref{treeuni3} by all the couplings.  As one can see in Figs.~\ref{ls-RGE}-\ref{rs-RGE}, close to $Q\approx10^{6.5}~\mathrm{GeV}$ the running couplings start to enter the regions of values which are forbidden by the Tree-Unitarity, see points in Figs.~\ref{ls-RGE}-\ref{rs-RGE} marked by black dots.

\noindent
If we choose higher values of initial parameters in Eq.~(\ref{IC}), e.g. increasing $M_{\rm H^{ FCNC}}$, Landau poles shift in the direction of lower $Q$ values.
It means that a region of stability decreases further, below $Q \sim 10^6$ GeV.

\section*{Conclusions and Outlook}
\noindent
The constraints from  Tree-Unitarity  give a good handle to understand the spectrum of the heavy scalar fields within the Left-Right symmetric models. We expressed TU in terms of the physical scalar fields, thus we have been able to  translate those constraints into the maximal mass limits of some beyond  Standard Model heavy particles. Along with that we impose the vacuum stability criteria 
to further constrain the parameter space. We have discussed the status of the benchmark points which we suggested in our earlier paper compatible with lack of FCNC effects, and which are interesting in the context of the LHC phenomenological aspects. All these constraints together leave a well defined
room in the parameter space, as a function of $v_R$. In the process we have come up with   general and complete set of 1-loop renormalization group equations for all couplings of the considered model.  We have performed evolutions of quartic couplings using these complete set of RGEs and shown how large the right-handed scale can be. It appears that restrictions coming from TU and RGEs  meet  approximately at the same $Q$ scale which  is also controlled by the choice of  $M_{\rm H^{FCNC}}$.

One of the possible future directions is to  perform the full 2-loop analysis of RG flow of scalar potential parameters taking into account the impact of the threshold corrections and proper matching conditions. They are crucial when one allows  large mass splitting among the heavy scalars.  And this feature is important for further phenomenological studies.
To ensure proper breaking of the electroweak symmetry, a bottom-up approach would be more appropriate for such analysis. 
Another important fact in this kind of analysis is a possible emergence of Landau poles at relatively low $Q$ scale which  signals  that the  perturbativity can be in trouble. Thus  one should perform this computation for higher orders as well with the hope that incorporation of 2-loop corrections may alleviate this problem.
 In this context the impact of heavy right-handed neutrino Yukawa coupling at the two-loop level cannot be ignored, and their roles have been already noticed at the 1-loop level low-energy muon decay analysis in \cite{Chakrabortty:2012pp}. Similar to the other fermion loops, the higher order corrections involving heavy neutrinos also contribute negatively to the beta functions of the quartic couplings and thus can delay the disaster of hitting the Landau pole. It is also known that 2-loop contributions to RGEs can significantly change running especially in the regime where parameters are bigger than $1$.

\section*{Acknowledgements}
\noindent
JG greatly appreciates the warm hospitality of Dept. of Physics, IIT Kanpur where part of the work was done.
We would like to thank Ira Rothstein for comments on RGEs of the MLRSM and Florian Lyonnet for comments and suggestions related to the usage of \textsc{PyR@TE} package.  Work   supported  by Department of Science \& Technology, Government of INDIA under the Grant Agreement number IFA12-PH-34 (INSPIRE Faculty Award) and by the Polish National Science Centre (NCN), Grant  No.~DEC-2013/11/B/ST2/04023.


\section*{Appendix}
The full scalar potential includes left and right-handed triplets \cite{Gunion:1989in,Deshpande:1990ip,Duka:1999uc}:
\begin{eqnarray}\label{V}
 & & V(\phi,\Delta_L,\Delta_R) =   \nonumber \\
    &+& \lambda_1\bigg\{\Big(\Tr\big[\phi^\dagger \phi\big]\Big)^2\bigg\} +
    \lambda_2\bigg\{ \Big(\Tr\big[\tilde{\phi}\phi^\dagger\big]\Big)^2+\Big(\Tr\big[\tilde{\phi}^\dagger \phi\big]\Big)^2 \bigg\} \nonumber\\
    &+& \lambda_3\bigg\{\Tr\big[\tilde{\phi}\phi^\dagger\big]\Tr\big[\tilde{\phi}^\dagger \phi\big] \bigg\} \nonumber  \\
    &+& \lambda_4 \bigg\{ \Tr\big[\phi^\dagger \phi\big]\Big(\Tr\big[\tilde{\phi}\phi^\dagger\big]
    +\Tr\big[\tilde{\phi}^\dagger \phi\big]\Big) \bigg\}\nonumber\\
    &+& \rho_1 \bigg\{ \Big(\Tr\big[\Delta_L \Delta_L^\dagger\big]\Big)^2+\Big(\Delta_R \Delta_R^\dagger\Big)^2 \bigg\} \nonumber \\
    &+& \rho_2 \bigg\{\Tr\big[\Delta_L \Delta_L\big]\;\Tr\big[\Delta_L^\dagger \Delta_L^\dagger\big]
    +\Tr\big[\Delta_R \Delta_R\big]\;\Tr\big[\Delta_R^\dagger \Delta_R^\dagger\big]  \bigg\}\nonumber\\
    &+& \rho_3 \bigg\{\Tr\big[\Delta_L\Delta_L^\dagger\big]\;\Tr\big[\Delta_R\Delta_R^\dagger\big]\bigg\} \nonumber \\
    &+& \rho_4  \bigg\{\Tr\big[\Delta_L\Delta_L \big]\;
    \Tr\big[\Delta_R^\dagger \Delta_R^\dagger\big] + \Tr\big[\Delta_L^\dagger \Delta_L^\dagger \big]\;
    \Tr\big[\Delta_R \Delta_R \big] \bigg\} \nonumber \\
  &  + & \alpha_1 \bigg\{\Tr\big[\phi^\dagger \phi\big]\Big(\Tr\big[\Delta_L\Delta_L^\dagger\big]
    +\Tr\big[\Delta_R\Delta_R^\dagger\big]\Big)\bigg\} \nonumber  \\
    &+& \alpha_2
    \bigg\{\Tr\big[\phi\tilde{\phi}^\dagger\big]\Tr\big[\Delta_R\Delta_R^\dagger\big] 
    + \Tr\big[\phi^\dagger\tilde{\phi}\big]\Tr\big[\Delta_L\Delta_L^\dagger\big]\bigg\} \nonumber \\
    &+& \alpha_2^*
    \bigg\{\Tr\big[\phi^\dagger\tilde{\phi}\big]\Tr\big[\Delta_R\Delta_R^\dagger\big] 
    + \Tr\big[\tilde{\phi}^\dagger\phi\big]\Tr\big[\Delta_L\Delta_L^\dagger\big]\bigg\} \nonumber  \\
    &+& \alpha_3 \bigg\{ \Tr\big[\phi \phi^\dagger \Delta_L \Delta_L^\dagger\big]
    +\Tr\big[\phi^\dagger \phi \Delta_R \Delta_R^\dagger\big]\bigg\}\nonumber\\    
    &-&\mu_1^2\Tr[\phi^\dag\phi]-\mu_2^2(\Tr[\widetilde{\phi}\phi^\dag]+\Tr[\widetilde{\phi}^\dag\phi])\nonumber\\
    &-&\mu_3^2(\Tr[\Delta_L\Delta_L^\dag]+\Tr[\Delta_R\Delta_R^\dag]).
\end{eqnarray}

After spontaneouss symmetry breaking of the above potential, the mass matrix which includes $M_{H_0^0}$ can be written in the following form
(for details, see \cite{Gunion:1989in})

\begin{equation}
M=\left(
\begin{array}{ccc}
 2 \epsilon ^2 \text{$\lambda_1$} & 2 \epsilon ^2 \text{$\lambda_4$} & \text{$\alpha_1$} \epsilon  \\
 2 \epsilon ^2 \text{$\lambda_4$} & \frac{1}{2} \left[4 (2 \text{$\lambda_2$}+\text{$\lambda_3$}) \epsilon ^2+\text{$\alpha_3$}\right] &
   2 \text{$\alpha_2$} \epsilon  \\
 \text{$\alpha_1$} \epsilon  & 2 \text{$\alpha_2$} \epsilon  & 2 \text{$\rho_1$} \\
\end{array}
\right).
\end{equation}

Expanding eigenvalues of this matrix in a small $\epsilon = \kappa_+/v_R$ parameter Eq.~(\ref{spec1}) emerges.

The benchmark point considered in the paper resulting in a mass spectrum obtained  with the following set of parameters ($v_R = 12\,\mathrm{TeV}$):
\begin{eqnarray}
&&\lambda_1 = 0.13,\;
\lambda_2 = 0,\;
\lambda_3 = 0,\;
\lambda_4 = 0,\\
&&\alpha_1 = 0, \;
\alpha_2 = 0, \;
\alpha_3 = 1.39, \\
&&\rho_1 = 1.0, \;
\rho_2 = 7.5\times10^{-4}, \;
\rho_3 = 2.003.
\end{eqnarray}  
All masses are given in GeV:
\begin{eqnarray}
&&M_{H^0_0} = 125,\\ 
&&M_{H^0_1}=  10000,\; 
M_{H^0_2} = 16971, \;
M_{H^0_3} = 465, \\
&&M_{A^0_1} = 10000,\; 
M_{A^0_2} = 465,\\
&&M_{H^\pm_1} = 487,\;\;\;
M_{H^\pm_2} = 10001,\\
&&M_{H_1^{\pm \pm}} = 508,\;\;\;
M_{H_2^{\pm \pm}} = 508.
\label{B2MH2pp}
\end{eqnarray}

Here, we present an example of 1-loop RGE generated with the help of \textsc{PyR@TE} (v1.2.2 beta) package:
\begin{eqnarray}
&&(4\pi)^2 \frac{d \lambda_1}{d\ln Q}=\nonumber\\
&&6\alpha_1^2 +6\alpha_1\alpha_3+\frac{5}{2}\alpha_3^2 +\frac{9}{8}g_{L}^4+\frac{3}{4}g_{L}^2g_{R}^2+\frac{9}{8}g_{R}^4\nonumber\\ 
&&-9g_{L}^2\lambda_1-9g_{R}^2\lambda_1+32\lambda_1^2+64\lambda_2^2\nonumber\\ 
&&+16\lambda_1\lambda_3 +16\lambda_3^2+48\lambda_4^2 +2\lambda_1\Tr\left(\widetilde{h}_l^\dag \widetilde{h}_l\right)\nonumber\\
&&+2\lambda_1\Tr\left(h_l^\dag h_l\right) +6\lambda_1\Tr\left(\widetilde{h}_q^\dag \widetilde{h}_q\right)+6\lambda_1\Tr\left(h_q^\dag h_q\right)\nonumber\\
&&+2\lambda_1\Tr\left(\widetilde{h}_l^* \widetilde{h}_l^T\right)+2\lambda_1\Tr\left(h_l^* h_l^T\right)+6\lambda_1\Tr\left(\widetilde{h}_q^* \widetilde{h}_q^T\right)\nonumber\\
&&+6\lambda_1\Tr\left(h_q^*h_q^T\right)-\Tr\left(\widetilde{h}_l \widetilde{h}_l^\dag \widetilde{h}_l \widetilde{h}_l^\dag\right)-\Tr\left(h_l h_l^\dag h_l h_l^\dag\right)\nonumber\\
&&-3\Tr\left(\widetilde{h}_q \widetilde{h}_q^\dag \widetilde{h}_q \widetilde{h}_q^\dag\right)-3\Tr\left(h_q h_q^\dag h_q h_q^\dag\right)\nonumber\\
&&-\Tr\left(\widetilde{h}_l^T \widetilde{h}_l^* \widetilde{h}_l^T \widetilde{h}_l^*\right)-\Tr\left(h_l^T h_l^* h_l^T h_l^*\right)\nonumber\\
&&-3\Tr\left(\widetilde{h}_q^T \widetilde{h}_q^* \widetilde{h}_q^T \widetilde{h}_q^*\right)-3\Tr\left(h_q^T h_q^* h_q^T h_q^*\right),
\end{eqnarray}
where $h_l$, $\widetilde{h}_l$, $h_{q}$ and $\widetilde{h}_q$ are Yukawa couplings, $\lambda_i$ and $\alpha_j$ are the scalar quartic couplings, $g_k$ are the gauge couplings as defined in Eq.~(14) in \cite{Duka:1999uc}.

\section*{References}
\providecommand{\href}[2]{#2}
\addcontentsline{toc}{section}{References}
\bibliography{LR_theory_constraints_arxiv_V2.bbl}

\end{document}